\documentclass{PoS}

\usepackage{epsfig}
\usepackage{graphicx}

\usepackage{axodraw}
\usepackage{pstricks}
\usepackage{amsmath}

\newcommand{\dbar}{\overline{d}}

\newcommand*{\CPT}{\raise0.4ex\hbox{$\chi$}PT}
\newcommand*{\chpt}{\raise0.4ex\hbox{$\chi$}PT}
\newcommand*{\schpt}{S\raise0.4ex\hbox{$\chi$}PT}

\def\eqref#1{{(\ref{#1})}}

\def\bar{\overline}

\title{Taste violations in the scalar correlator in mixed action simulations}

\ShortTitle{Taste violations in the scalar correlator in mixed action simulations}

\author{\speaker{C.~Aubin}\\
Dept.\ of Physics,
Columbia University, 
New York, NY, USA\\
Dept.\ of Physics,
College of William \& Mary, 
Williamsburg, VA, USA\thanks{Present address.}\\
E-mail: \email{caaubin@wm.edu}}
\author{Jack Laiho\\
Theoretical Physics Department, 
Fermi National Accelerator Laboratory, Batavia, IL, USA\thanks{Operated by Fermi  Research Alliance, LLC,
         under Contract No.~DE-AC02-07CH11359 with the  United States Department of Energy.}\\
Dept.\ of Physics,
Washington University, St.\ Louis, MO, 
USA\thanks{Present address.}\\
E-mail: \email{jlaiho@fnal.gov}}
\author{Ruth S.\ Van de Water\\
Theoretical Physics Department, 
Fermi National Accelerator Laboratory, Batavia, IL, USA\\
E-mail: \email{ruthv@fnal.gov}
}

\abstract{We study the behavior of the isovector scalar correlator, which is particularly sensitive to lattice artifacts, using domain-wall valence quarks on a staggered sea (generated by the MILC collaboration). 
 We analyze this according to the prediction from chiral perturbation theory determined by Prelovsek, which indicates that the leading unitarity violations come from taste breaking effects. We show that our data behaves in the way predicted by Prelovsek, thus verifying that the largest contribution to the violations of unitarity which arise at finite lattice spacing can be described by the mixed-action chiral perturbation theory.}

\FullConference{The XXV International Symposium on Lattice Field Theory\\
		 July 30-4 August 2007\\
		 Regensburg, Germany}

\makeatletter\let\default@color\current@color\makeatother

\begin{document}

\section{Introduction}

Currently, an increasing number of mixed-action simulations are being performed by various groups (for example, see Refs.~\cite{Bistrovic:2005fd,Beane:2006gj,Beane:2006mx,ruthtalk}) to measure various quantities of interest. There are several reasons why this is a useful endeavor, both practical and theoretical. Practically speaking, one can, for example, use existing gauge field configurations to calculate the quantities one is interested in, but not be restricted to the same discretization of quarks. To simplify the analysis of lattice data, the symmetries of the valence sector are much more important to retain, so one should choose a valence discretization which has the symmetries one desires most. Additionally, there is both a theoretical and practical interest in using different methods to obtain the same physical quantity. While different calculations with different discretizations are useful, one can add more cross checks by mixing various discretizations for the same quantities.

A theoretical issue which arises in mixed action simulations is that unitarity is violated at non-zero lattice spacing \cite{Golterman:2005xa}.  This is due to the fact that at non-zero lattice spacing the valence and sea sectors have different discretization effects so that any tuning is only exact up to lattice spacing dependent terms.  Thus a mixed action theory is necessarily partially quenched, and this violation of unitarity is much more pronounced and only goes away in both the continuum limit \emph{and} the limit where the sea and valence masses become equal. One cannot reach the full QCD limit, where $m_{\rm sea} = m_{\rm val}$, at finite lattice spacing.

We are currently calculating the kaon B-parameter using a mixed action \cite{ruthtalk,Aubin:2006hg}, and the question is: Can we theoretically understand, for a given quantity, the primary source of the violation of unitarity, and thus remove this artifact to reveal the physical quantities? For many quantities, such as $m_\pi$ or $B_K$, for example, the violation of unitarity shows up mildly in the chiral expressions, and one would like to see a more pronounced violation of unitarity that can still be understood using chiral perturbation theory (\chpt). In this work we show that this can in fact be done when analyzing the isovector-scalar correlator, the $a_0$, using mixed-action \chpt\ \cite{Bardeen:2001jm,Prelovsek:2005rf}.
Quantities such as the $a_0$ correlator (as well as $\pi-\pi$ scattering in the $I=0$ channel) are particularly sensitive to this effect, due to flavor-neutral intermediate states. These include disconnected (at the quark level) diagrams, where unitarity violations are more pronounced, and for the $a_0$ specifically, this affects the lattice correlator itself.
While the issues we discuss are generic to all mixed-action simulations, we will focus on those where the sea quarks are staggered and the valence quarks are domain wall.

\section{Mixed-Action \chpt}

The formalism of \chpt\ for a mixed action was laid down initially in Ref.~\cite{Bar:2005tu} in the context of $m_\pi,f_\pi$,
and has been extended for many other quantities. Although we will not describe mixed-action \chpt\ in detail, we would like to look at the three different types of mesons that can arise: those made from two sea quarks, those from two valence quarks, and finally mesons with one of each type of quark. For the two valence quarks, we have the tree-level relation for the masses
\begin{equation}
	m_{vv'}^2 = \mu(m_v + m_{v'} + 2m_{\rm res})
\end{equation}
where $m_{\rm res}$ is the residual mass\footnote{Recall that the residual mass arises from domain-wall quarks by the overlap of the left- and right-handed modes with a finite-sized fifth dimension. $m_{\rm res}$ is a measure of the chiral symmetry breaking caused by this overlap.} and $v,v'$ are two valence quark flavors.
With two staggered sea quarks, a meson of a given flavor also has associated with it a taste $t$ \cite{Lee:1999zx,Aubin:2003mg}
\begin{equation}
	m_{ss'}^2 = \mu(m_s + m_{s'}) + a^2 \Delta_t\ ,
\end{equation}
where $t$ runs over the 16 tastes which fall into five multiplets transforming in irreducible representations of the remaining $SO(4)$ taste symmetry: $t\in\{P,A,T,V,I\}$ \cite{Lee:1999zx,Aubin:2003mg}. For the pseudoscalar taste, $\Delta_P = 0$, which is a manifestation of the fact that this staggered meson is a Goldstone boson at finite lattice spacing when $m_{s,s'}\to0$. The relevant taste splitting in mixed-action simulations is the singlet taste, $\Delta_I$ (the valence quarks, being taste singlets, couple only to the taste-singlet meson in the sea sector) and this is also the largest of all the taste splittings. Finally, a meson with one sea and one valence quark has the mass
\begin{equation}
	m_{vs}^2 = \mu(m_v + m_s + m_{\rm res}) + a^2\Delta_{\rm mix}\ ,
\end{equation}
where $\Delta_{\rm mix}$ is a new splitting, unique to the particular choice of mixed action, that arises from four-quark operators in the same way as the staggered taste splittings arise \cite{Bar:2005tu}. Although this parameter does not appear in many mixed action quantities (such as $m_{\pi}^2$ or $B_K$), it will play an important role in $f_K$ and also the scalar correlator.

\section{The $a_0$ in mixed-action \chpt}

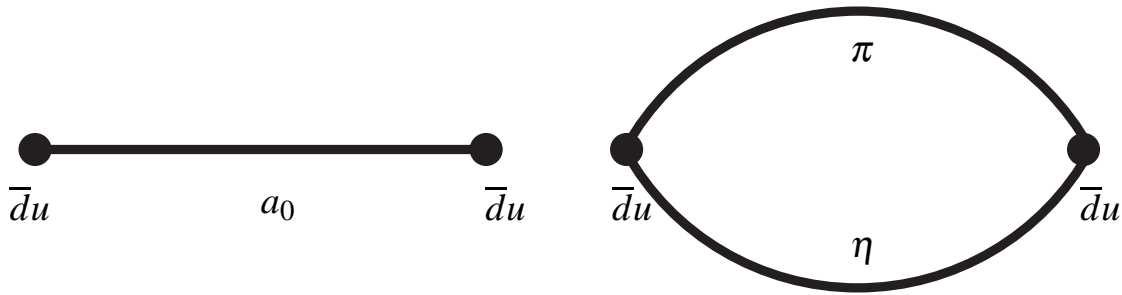
\begin{figure}[t]
\begin{center}
  \begin{picture}(200,110)(0,-10)
    \SetWidth{3.5}
    \Line(-100.0,27.62)(70.0,27.62)
    \Text(-110.0,0.89)[lb]{\Large{$\bar d u$}}
    \Text(70.0,0.89)[lb]{\Large{$\bar d u$}}
    \Text(-15.0,0.89)[lb]{\Large{$a_0$}}
    \SetWidth{0.5}
    \Vertex(-100.0,27.62){6.24}
    \Vertex(70.0,27.62){6.24}
    \SetWidth{3.5}
    \CArc(210.0,-20.19)(100.0,28.94,151.06)
    \CArc(210.0,75.43)(100.0,-151.06,-28.94)
    \SetWidth{0.5}
    \Vertex(122.93,27.62){6.24}
    \Vertex(295.12,27.62){6.24}
    \Text(117.93,0.89)[lb]{\Large{$\bar d u$}}
    \Text(295.12,0.89)[lb]{\Large{$\bar d u$}}
    \Text(207.85,60.81)[lb]{\Large{$\pi$}}
    \Text(207.85,-15.93)[lb]{\Large{$\eta$}}
  \end{picture}
\end{center}
\label{fig:bubble}
\caption{The two leading terms from the scalar current: The first is the direct term, corresponding to the propagation of an $a_0$ meson, while the second is one possible ``bubble'' term, where there is a $\pi$ and an $\eta$ propagating (we can also have $\pi-\pi$ and $K-\overline{K}$ intermediate states as well).}
\end{figure}

The isovector scalar is created using the following local current at the underlying quark level and the chiral level (since we will be using \chpt\ to calculate the contribution to the scalar correlator coming from two-particle intermediate states):
\begin{equation}
	S(\mathbf{x},t) = \dbar(\mathbf{x},t) u(\mathbf{x},t)\ ,
	\qquad
	S_\chi(x) = \mu \left[\Phi^2(x)\right]_{ud}\ .
\end{equation}
We are interested in the lattice correlator given by $C(t) = \sum_{\mathbf{x}}\left\langle 0|S(\mathbf{x},t)S^\dag(\mathbf{0},0)|0\right\rangle$.
 The leading term in this correlator corresponds to the propagation of an $a_0$ from time 0 to time $t$. However, as was first noticed in the quenched case \cite{Bardeen:2001jm}, there is a sizable contribution to the correlator coming from two-particle intermediate states, shown in Fig.~\ref{fig:bubble}. Thus, instead of using the usual single-exponential expression to fit the lattice correlator, we use the form
\begin{equation}\label{eq:corr}
	C(t) = A e^{-m_{a_0} t } + B(t) + \cdots
\end{equation} 
where the $\cdots$ represent excited state contributions that we will neglect. The ``bubble term'' $B(t)$ has been calculated by Prelovsek \cite{Prelovsek:2005rf} using mixed-action \chpt, with the result for 2+1 flavors of sea quarks (taking the time direction to be infinite in length)
\begin{eqnarray}\label{eq:bubble}
	B(t)
	&=& \frac{\mu^2}{3L^3}\sum_{\mathbf{k}}
	\Biggl[
	\frac{2}{9}\frac{e^{-(\omega_{vv} + \omega_{\eta_I}) t}}
	{\omega_{vv}\omega_{\eta_I}}
	\frac{(m_{S_I}^2 - m_{U_I}^2)^2}
	{(m^2_{vv} - m^2_{\eta_I})^2}
	-\frac{e^{-2\omega_{vv} t}}{\omega_{vv}^2}
	\left[
	\frac{3m^2_{vv}(m^2_{vv} - 2m^2_{\eta_I})
	+ 2m^4_{S_I} + m^4_{U_I}}
	{3(m^2_{\eta_I} - m^2_{vv})^2 }\right]
	\nonumber\\&&{}
	- \frac{e^{-2\omega_{vv} t}}{2\omega_{vv}^4}
	\left(\omega_{vv} t + 1\right)
	\frac{(m^2_{U_I} - m^2_{vv})
	(m^2_{S_I} - m^2_{vv} )}{m^2_{\eta_I} - m^2_{vv}}
	+ \frac{3}{2}\frac{e^{-2\omega_{vu} t}}{\omega_{vu}^2}
	+ \frac{3}{4}\frac{e^{-2\omega_{vs} t}}{\omega_{vs}^2}
	\Biggr]\ ,
\end{eqnarray}
where we have defined $\omega_i^2 = \sqrt{\mathbf{k}^2 + m_i^2}$, and $m_{\eta_I}^2 = (m_{U_I}^2 + 2m_{S_I}^2)/3$. 

There are several interesting things to note here. Once $\Delta_{\rm mix}$ and $\Delta_I$ (which appear in $m_{vu}, m_{vs},$ and $m_{U_I},m_{S_I}$) are known, then there are no free parameters in this expression. The meson masses can all be measured and the coefficient $\mu$ can be determined as well from spectrum calculations.\footnote{It is interesting to also note that one does not actually need to have measured the residual mass, because if one merely inputs the value for $m_{vv}$ as measured for a given valence quark mass $m_v$, then $m_{\rm res}$ is implicitly included in the meson mass.} Both the shape and the normalization are completely predicted by the \chpt.

Additionally, there is a negative residue in the momentum-space bubble term, which leads to the two negative terms above. The third term is especially dangerous because of the linear-in-$t$ growth factor in front of the exponential. This term dominates at intermediates times, and if the masses are light enough, as we will show, this will cause the correlator to become negative. This linear rise comes about by the existence of a double pole in the momentum-space correlator, and we can get rid of this term with an appropriate tuning of the masses, as discussed next, but this will not fully solve our problem.

This occurs because this theory is not unitary at finite lattice spacing. In fact, one could tune the valence pion masses to the sea pion masses to try to obtain ``full QCD,'' but no unique point exists at finite lattice spacing. This can be seen especially in this case: The ``natural'' tuning would be to set $m_{vv} = m_{U_P}$, since these both will vanish in the chiral limit at finite lattice spacing. However, this only makes the troublesome term in $B(t)$ larger, and if we were to try to get rid of this double pole, we should set $m_{vv} = m_{U_I}$, and the $\omega_{vv}t$ term in $B(t)$ vanishes.\footnote{On the coarse MILC lattices this may not be an advisable tuning, since then we will have a rather heavy valence pion.} Note though, that this does not guarantee positivity, since there still is another possible negative term in the bubble (it is just no longer enhanced by the double pole). Additionally, although we could in principle tune the theory to fix the enhancement of the unitarity violation in the $a_0$ correlator (and other quantities such as the $I=0$ $\pi-\pi$ scattering phase shift, for example \cite{Golterman:2005xa}), it will still violate unitarity to some degree. We cannot completely solve this problem at finite lattice spacing.

The key here is that there is no such concept as full QCD for a mixed-action (any tuning one chooses will still never get rid of the negative residue in the scalar correlator). It is true that this violation vanishes in the continuum limit, so full QCD is recovered, since we expect both rooted-staggered quarks and domain-wall quarks to reproduce the same continuum theory as $a\to 0$.\footnote{Although there is no rigorous proof, there is considerable supporting evidence that the rooting procedure for staggered fermions is correct. We work under the plausible assumption that using the rooting procedure we recover QCD in the continuum limit (see Ref.~\cite{Sharpe:2006re} and references therein).}

We have measured the meson masses with two valence quarks as well as $\Delta_{\rm mix}$, and for a detailed discussion of this, see \cite{ruthtalk}. We have presented these results previously in \cite{Jacktalk} and our results agree with an independent determination in Ref.~\cite{Orginos:2007tw}. Additionally, we can use the values of $\Delta_I$ that have been determined by the MILC collaboration \cite{Aubin:2004wf} to input into the bubble formula.

\section{Lattice results for scalar correlator}

\begin{figure}[ht]
\begin{center}
	\epsfxsize=3in \epsffile{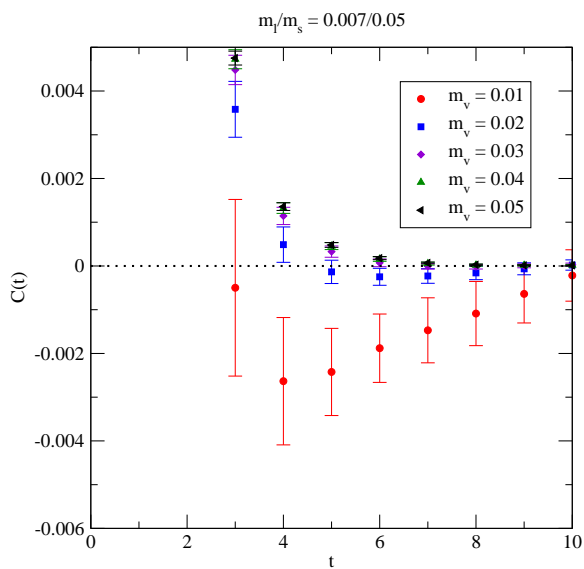} 
\caption{Scalar correlator data for various valence masses and a sea mass of $0.007$.}\label{fig:alldata}
 \end{center}
\end{figure}

In Fig.~\ref{fig:alldata} we show a portion of the data for the scalar correlator that we have accumulated thus far. The data shown is for the coarse ($a\sim 0.12$ fm) MILC ensemble with a light sea quark mass of $0.007$, with $am_v \in \{0.01, 0.02, 0.03$, $0.04, 0.05\}$. 
Fig.~\ref{fig:alldata} shows the qualitatively expected behavior based on the expression for B(t) in Eq.~(\ref{eq:bubble}).  In particular, the size of the negative bubble contribution decreases as the valence quark mass increases, such that the correlator stays positive for all times in the data.

\begin{figure}[t]
\begin{tabular}{cc}
    \epsfxsize=2.7in \epsffile{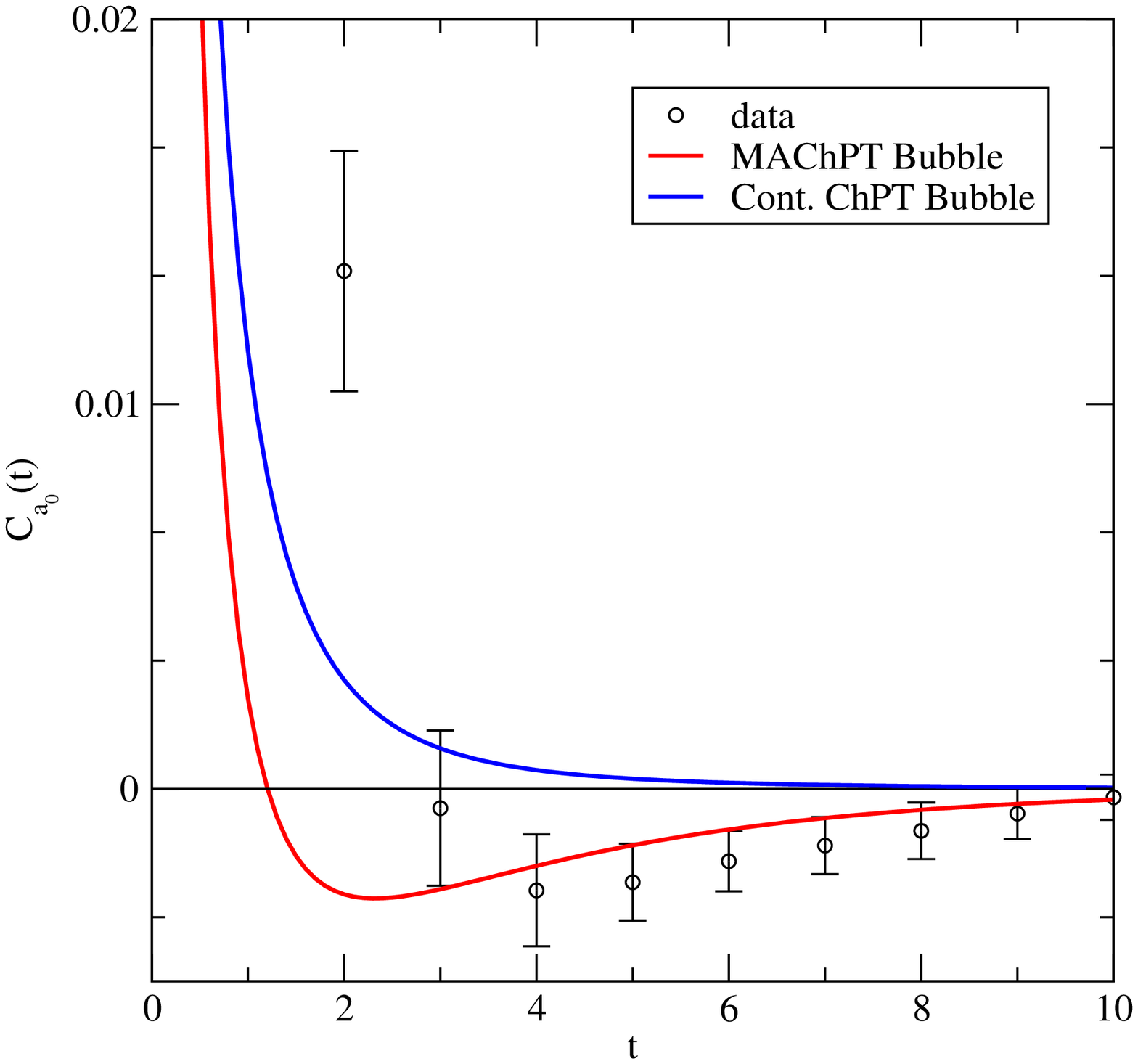} & 
    \epsfxsize=2.7in \epsffile{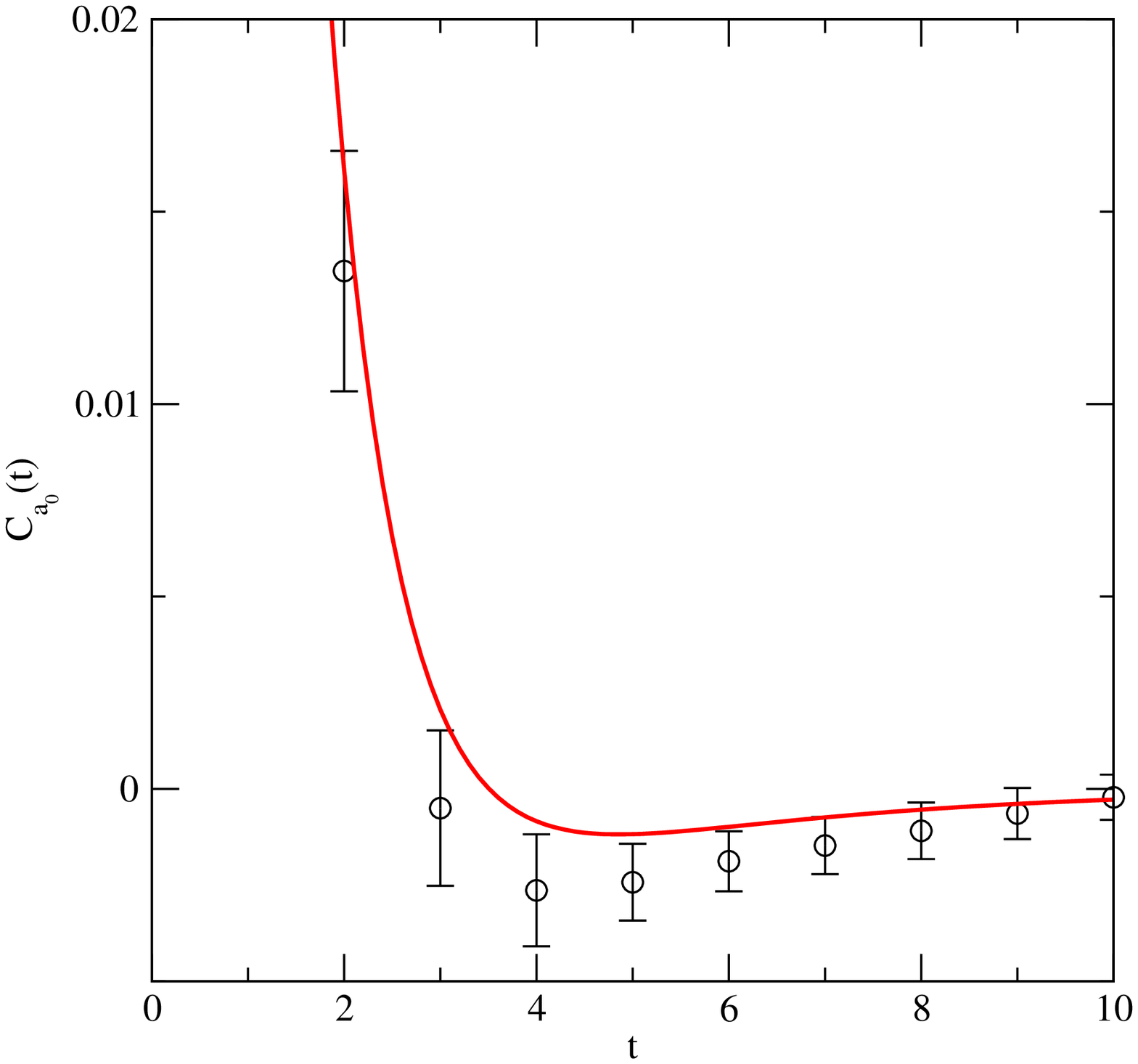}  \\
    (a) & (b)  \\
    \end{tabular}
    \caption{(a) Scalar correlator data for $m_{\rm val} = 0.01$ and $m_{\rm sea} = 0.007$ (data points) and the bubble function with (solid red line) and without (solid blue line) the splittings included. (b) Scalar correlator data for $m_{\rm val} = 0.01$ and $m_{\rm sea} = 0.007$, with a fit (red line) to $C(t)$.}
    \label{fig:fits}
\end{figure}

In Fig.~\ref{fig:fits}(a) we show the prediction for the bubble contribution, which has no free parameters, overlaid on the scalar correlator data for the 0.007 sea quark mass and the 0.01 valence quark mass. The red line is the bubble function using the parameters for $\Delta_{\rm mix}$ as shown in \cite{ruthtalk}, and $\Delta_I$ as calculated by MILC \cite{Aubin:2004wf}. This is the bubble as predicted by mixed-action \chpt. The blue line is the same function but for the continuum limit, where we have set $\Delta_{\rm mix} = \Delta_I = 0$. We can see that for this set of masses, the continuum bubble does not go negative, and clearly cannot describe our data. This shows that it is necessary to use the correct \chpt\ formulation corresponding to the lattice action one chooses. Of course, for small times, the direct term $Ae^{-m_{a_0}t}$ dominates, so we wouldn't expect the bubble to match the data precisely, but for $t\ge3$, the mixed-action bubble is qualitatively consistent with our data.

Finally, in Fig.~\ref{fig:fits}(b), we show the data and a fit to the expression given in Eq.~(\ref{eq:corr}). Various fitting ranges give similar results with comparable correlated $\chi^2/d.o.f.$ (roughly 1.2 for the fit shown), yet the values for the $a_0$ mass in the exponential term vary greatly. Without a finer resolution in the time direction, we cannot hope to determine the $a_0$ mass with this expression. However, this is able to serve as a test of the methodology: The use of mixed-action \chpt\ seems to describe the low-energy behavior of our theory qualitatively quite well.\footnote{There are other possible violations of unitarity that we cannot resolve in our data and cannot be described by \chpt. These may be due to the non-positive-definite transfer matrix for domain-wall fermions \cite{Furman:1994ky}, and also enhanced zero-mode contributions (in the quenched case, this was studied in Ref.~\cite{Blum:2000kn}, for domain-wall fermions), which would be noticeable at small times were we to resolve our data more precisely.}

\section{Conclusion}

We see that the behavior of the scalar correlator is qualitatively predicted by mixed-action \chpt. This gives us confidence that although a mixed-action simulation is not unitary at finite lattice spacing, we can understand theoretically the primary source of this unitarity violation is, and more importantly can quantitatively account for it. Of course, with better statistics it might turn out that there are higher-order terms that come into play, and thus we will need a more sophisticated analysis. However, at this level, the negativity of the $a_0$ correlator is accounted for solely from the inclusion of a single, well-defined, bubble term, coming from two-particle intermediate states.

\section*{Acknowledgments}

We would like to thank the MILC Collaboration for making their configurations available for this study, as well as the USQCD Collaboration for allocating computer time. Additionally, we would like to thank Sasa Prelovsek for useful discussions.


\providecommand{\href}[2]{#2}\begingroup\raggedright\endgroup

\end{document}